# A kilobyte rewritable atomic memory

F. E. Kalff[1], M. P. Rebergen[1], E. Fahrenfort[1], J. Girovsky[1], R. Toskovic[1], J. L. Lado[2], J. Fernández-Rossier[2,3] and A. F. Otte[1]*

[1]Department of Quantum Nanoscience, Kavli Institute of Nanoscience, Delft University of Technology, Lorentzweg 1, 2628 CJ Delft, The Netherlands.

[2]International Iberian Nanotechnology Laboratory (INL), Avenida Mestre José Veiga, 4715-310 Braga, Portugal.

[3]Departamento de Física Aplicada, Universidad de Alicante, San Vicente del Raspeig, 03690 Spain.

*To whom correspondence should be addressed. E-mail: a.f.otte@tudelft.nl

**The advent of devices based on single dopants, such as the single atom transistor[1], the single spin magnetometer[2,3] and the single atom memory[4], motivates the quest for strategies that permit to control matter with atomic precision. Manipulation of individual atoms by means of low-temperature scanning tunnelling microscopy[5] provides ways to store data in atoms, encoded either into their charge state[6,7], magnetization state[8–10] or lattice position[11]. A defining challenge at this stage is the controlled integration of these individual functional atoms into extended, scalable atomic circuits. Here we present a robust digital atomic scale memory of up to 1 kilobyte (8,000 bits) using an array of individual surface vacancies in a chlorine terminated Cu(100) surface. The memory can be read and rewritten automatically by means of atomic scale markers, and offers an areal density of 502 Terabits per square inch, outperforming state-of-the-art hard disk drives by three orders of magnitude. Furthermore, the chlorine vacancies are found to be stable at temperatures up to 77 K, offering prospects for expanding large-scale atomic assembly towards ambient conditions.**

Since the first demonstration of atom manipulation, 25 years ago[5], the preferred approach for assembling atomic arrangements has been the lateral positioning of atoms or molecules evaporated onto a flat metal surface, most notably the (111) crystal surface of copper[12–15]. While ideal for experiments comprising up to several hundreds of constituents, the absence of a large-scale defect-free detectable grid on this surface prohibits the construction of architectures involving correlated lattice-placement of atoms separated by more than a few nanometres. Moreover, thermal motion of the adatoms restricts the technique to temperatures below 10 K. As we demonstrate below, we find that manipulation of missing atoms in a surface (vacancies)[16], as opposed to additional atoms atop, permits a dramatic leap forward in our capability to build functional devices on the atomic scale.

To this purpose, we take advantage of the self-assembly of chlorine atoms on the Cu(100) surface[17–20], forming a flat two-dimensional lattice with several convenient properties. First, it provides large areas of a perfect template grid, with a controllable coverage of vacancies. Second, the chlorine lattice remains stable up to a large density of vacancies and up to relatively high temperature (77 K). And third, critical for our purpose, the precise location of the vacancies can be manipulated by STM with a very high level of control (and without the need to pick-up atoms with the tip, i.e. vertical atom manipulation). As we show below, these properties allow us to position thousands of vacancies at predefined atomic sites in a reasonable timeframe.

The chlorinated copper surface is prepared in ultra-high vacuum through the evaporation of anhydrous $CuCl_2$ powder heated to 300 °C onto a clean Cu(100) crystal surface. The crystal is pre-heated to 100-150 °C prior the $CuCl_2$ deposition for approx. 12 minutes and kept at this temperature during the deposition and the 10 minutes post-anneal. This results in the formation of a square c(2×2) reconstruction of Cl atoms with a lattice constant $a$ of 0.36 nm. The Cl coverage (and thereby the vacancy coverage $x$) can be tuned by varying the duration of evaporation. E.g., an evaporation time of 240 s led to a vacancy coverage of $x = 0.169$, determined using STM, whereas a time of 210 s gave $x = 0.115$.

When imaged by STM, vacancies are resolved as square depressions ~20-30 pm deep (Fig. 1). According to density functional theory (DFT) calculations (see Supplementary Information for details), a Cl atom has to overcome an energy barrier $\Delta = 0.3$ eV in order to swap places with a neighbouring vacancy (Figs. 1a and b). As a result, the surface is resilient to tunnelling currents of up to 2 µA when imaged at positive sample voltages of ~200 mV or lower.

As shown in Figs. 1c and d, vacancies can be moved by injecting a current of $1.0 \pm 0.5$ µA (error represents variations in the exact shape of the STM tip apex) at +500 mV sample voltage at a position approximately $0.4a$ along the way from the centre of the vacancy to the centre of the neighbouring Cl atom at the desired location. The STM feedback is kept switched on throughout the manipulation procedure. Although several attempts may be required to make a vacancy move, the directional reliability (i.e. how often a vacancy moves in the desired direction once it moves) can be in excess of 99%, depending on tip shape. Controlled vacancy movement is limited to the (±1,0) and (0,±1) directions on the square Cl grid. Diagonal moves (e.g., in the (1,1) direction) were found to occur sporadically, but could not be induced controllably.

In our structure, we define a vertical pair of a chlorine atom (Cl) and a vacancy (V) as a bit, where the V-Cl configuration represents the "0" and Cl-V the "1". In order to avoid vacancies directly neighbouring each other, which would render automated locking of the STM tip on individual vacancies impossible, we implement a row of Cl atoms to separate bits in both the horizontal and vertical directions. For this reason, 6 lattice sites are needed for a bit, or 48 lattice sites for a byte (Fig. 2a), resulting in an optimal vacancy coverage $x = 1/6 = 0.167$.

We have not been able to controllably create or destroy vacancies without altering the tip apex. As a result, the vacancy coverage can be controlled only during sample preparation and cannot be changed afterwards. The bit arrays presented here were all made on a surface with $x = 0.115$. In order to compensate for the vacancy deficiency, the memory is organized in blocks of 8 bytes (64 bits) as shown in Figs. 2b-d, separated by a margin of 4 unit cells. Constructing this block configuration requires a vacancy coverage of $x = 0.118$, which is within 3% of the actual coverage. These blocks form a convenient way to organize the data. A read-out and a rewrite of a block take approximately 1 and 2 minutes, respectively.

We make use of an autonomous manipulation method[21] that permits the construction of large memories. A marker at the top left of each block defines the scan frame and the lattice for the complete block. After scanning the area, the positions of all vacancies are determined through image recognition. Next, a pathfinding algorithm is used to calculate the building sequence, guiding the vacancies to their respective final positions. The markers for adjacent blocks are built automatically as part of the construction and leftover vacancies are swept to the side to be used in future blocks. Automated construction of a complete block takes in the order of 10 minutes (see Supplementary Movie S1).

The scalability of the technique is demonstrated in Fig. 3 and Supplementary Figure S1, which show a complete memory consisting of 1016 bytes (8128 atomic bits) written to two different texts. Due to local defects or contaminants, some areas are not suitable for building switchable bits. Such local imperfections do not need to affect the functionality of the memory. By properly defining markers consisting of several vacancies, blocks can be designated as broken and will be skipped in the reading and writing sequences. Additional markers, denoting e.g. the start or ending of a line, allow for fully automated navigation of the STM tip through the memory. More complex markers may be designed to allow travel of the STM tip over longer distances and in arbitrary directions. Due to 12% of blocks being not suitable for data storage, the actual areal density of the complete memory comes down to 0.778 bits nm$^{-2}$, or 502 Terabits per square inch.

Our DFT calculations permit an order of magnitude estimate of the single vacancy switching rates. The estimated attempt frequency for a Cl atom to overcome the energy barrier $\Delta = 0.3$ eV, obtained for the small oscillation analysis around the equilibrium position, yields $\Gamma_0 = 48$ THz. Thus, we can estimate the thermally activated switching rate $\Gamma_0 \exp(-\Delta/k_B T)$, where $k_B$ is the Boltzmann constant, to be in the range of $10^{-5}$ Hz (several hours lifetime) at a temperature $T = 77$ K, although large error bars have to be assigned due to the exponential dependence on $\Delta$. As shown in Figs. 4a and b, taken 44 hours apart at 77.5 K, the manipulated vacancies are experimentally found to be stable for at least that amount of time, provided they are properly positioned. Vacancies arranged within two unit cells from each other are found to settle into a lower energy state within tens of minutes at this temperature (see Supplementary Figure S2).

In order to understand the vacancy-vacancy interaction, we have computed several configurations with two vacancies (Fig. 4c) in a supercell of up to 5×5 Cl atoms. We find that the minimal energy configurations for a pair is the diagonal dimer, i.e., two consecutive vacancies along the (1,±1) direction, which explains the observed high natural abundance of these. This finding goes a long way to account for the formation of a stripe phase with coexisting domains along the (1, ±1) directions, observed near $x = 1/3$ (Fig. 4d). Monte Carlo simulations (Fig. 4e), using the vacancy-vacancy interactions discussed above, are in very good agreement with the experiment and show that this system provides an ideal physical realization of a lattice gas model, for which stripe phases are expected[22].

Having several thousands of single-atom bits represents a tremendous step forward in the field of atomic scale electronics. There is no physical limitation that prevents the fabrication of much larger atomic memories. Making use of demonstrated high-frequency STM electronics[23], readout speeds in the order of 1 Mb/s should be attainable. While the technology presented here is two-dimensional, we foresee – taking advantage of the techniques to yield vertical stacking of 2D crystals[24] – a three-dimensional scaling-up of petabyte atomic memories, that would afford to store the US Library of Congress in a cube 100 μm wide, assuming a modest vertical pitch of 5 nm and the same in-plane pitch demonstrated here.

**Acknowledgments:**

We thank A. J. Heinrich for discussions. This work was supported by the Netherlands Organisation for Scientific Research (NWO/OCW), the Foundation for Fundamental Research on Matter (FOM), and by the Kavli Foundation. JFR and JLL acknowledge financial support by Marie-Curie-ITN Grant no. 607904-SPINOGRAPH. JFR acknowledges financial support by MEC-Spain (Grant no. FIS2013-47328-C2-2-P) and Generalitat Valenciana (PROMETEO 2012/011). For the subject-matter of this manuscript a Dutch patent application has been filed (Ref. no. NL2016335).

**Author contributions:**

FEK and EF developed the vacancy movement procedure; MPR, FEK and AFO programmed the autonomous vacancy manipulation; JG, MPR and RT performed the measurements at 77 K; JLL and JFR performed the DFT and Monte Carlo calculations; AFO devised the experiment and supervised the research. All authors discussed the results and contributed to writing the manuscript.

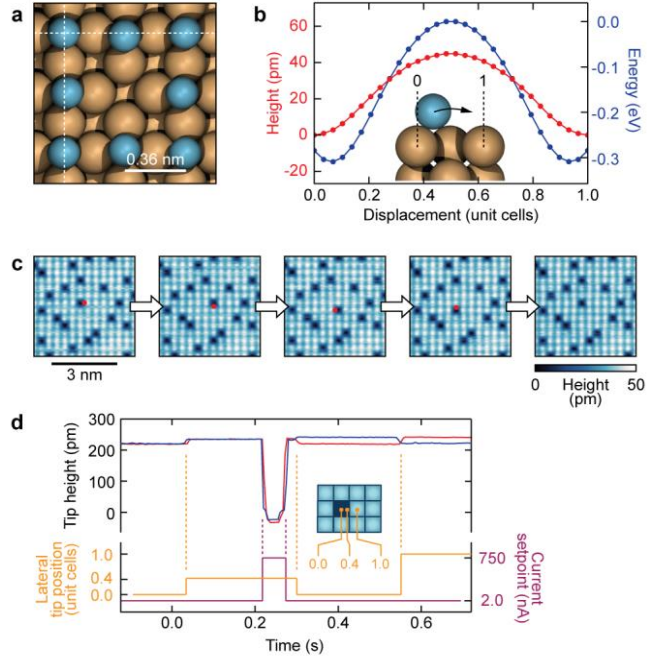

**Fig. 1. Hopping mechanism of a Cl vacancy on chlorinated Cu(100). a.** Atomic structure around a vacancy, calculated through DFT. Cl atoms (blue) are found to relax slightly towards the centre of the vacancy. **b.** DFT calculations showing the height profile and the corresponding potential energy of a Cl atom during a switch. **c.** Subsequent STM images (tunnelling current $I = 2.00$ nA, sample bias voltage $V = +500$ mV, temperature $T = 1.5$ K) of a vacancy being hopped in all four directions. The tip position for each manipulation is designated by a red dot. **d.** Measured tip height during a successful (blue) and an unsuccessful manipulation (red). Yellow and purple curves show the lateral tip position (see inset) and applied tunnelling current setpoint respectively. $V = +500$ mV throughout the manipulation. After the manipulation, the tip visits the original location of the Cl atom and the target position. Variations in the tip height between these two positions tell if the manipulation was successful.

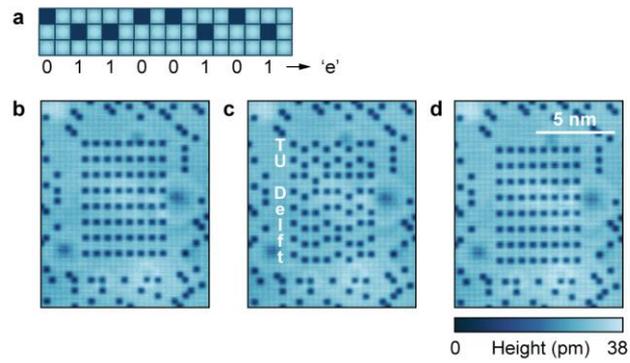

**Fig. 2. Data encoding principle. a.** Diagram showing the smallest possible byte, provided that vacancies may not be direct neighbours. The byte is set to represent the binary ASCII code for the character 'e'. **b–d.** STM images ($I$ = 2.00 nA, $V$ = +500 mV, $T$ = 1.5 K) of a 64-bit block, written as all 0's (b), a text (c), and all 1's (d).

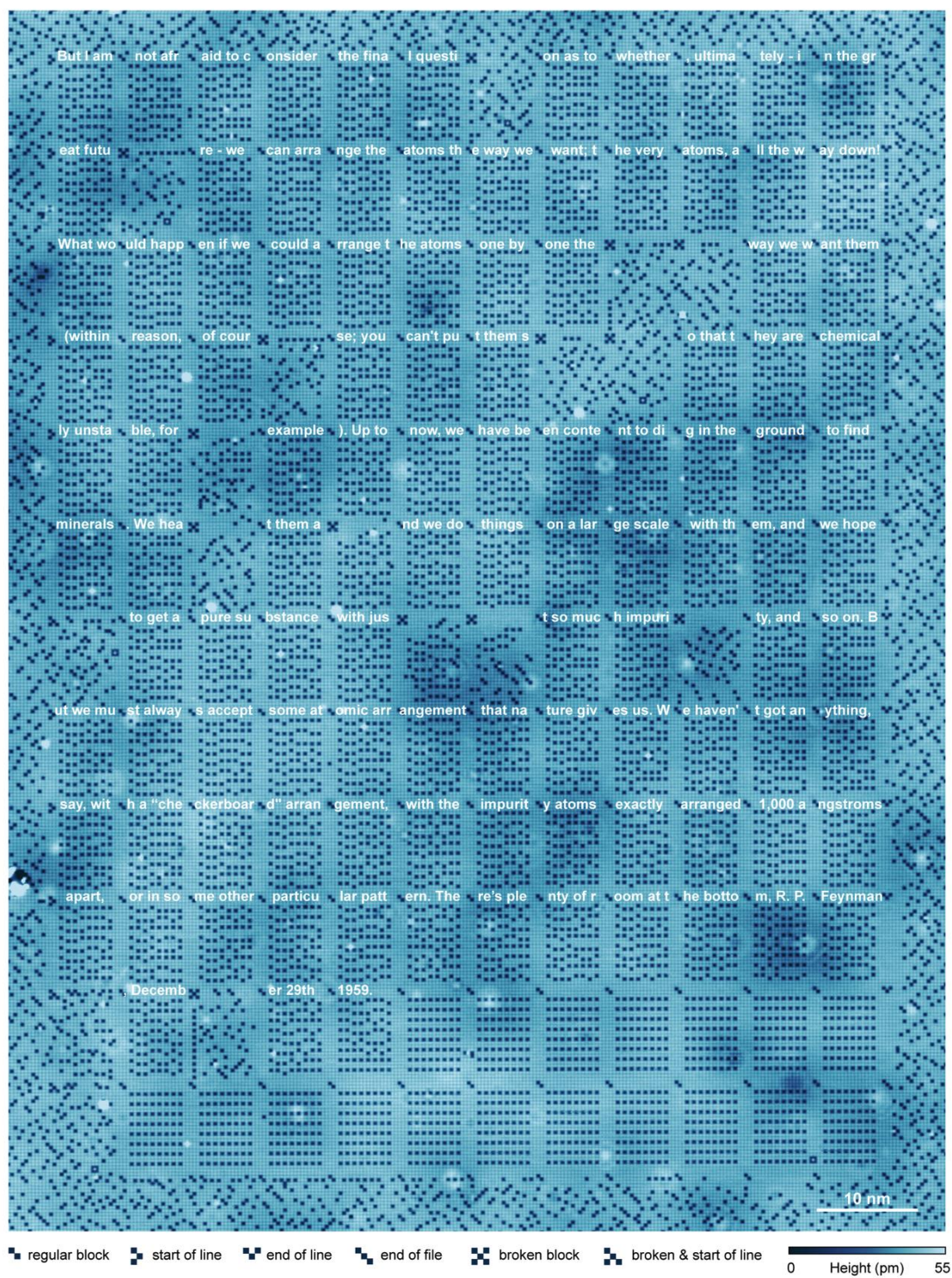

**Fig. 3. Kilobyte atomic memory.** STM image (96 nm × 126 nm, $I$ = 2.00 nA, $V$ = +500 mV, $T$ = 1.5 K) of a 1016 byte atomic memory, written to a passage from Feynman's lecture *There's plenty of room at the bottom*[25]. The various markers used are explained in the legend below. The memory consists of 127 functional blocks and 17 broken blocks, resulting in an overall areal density of 0.778 bits nm$^{-2}$.

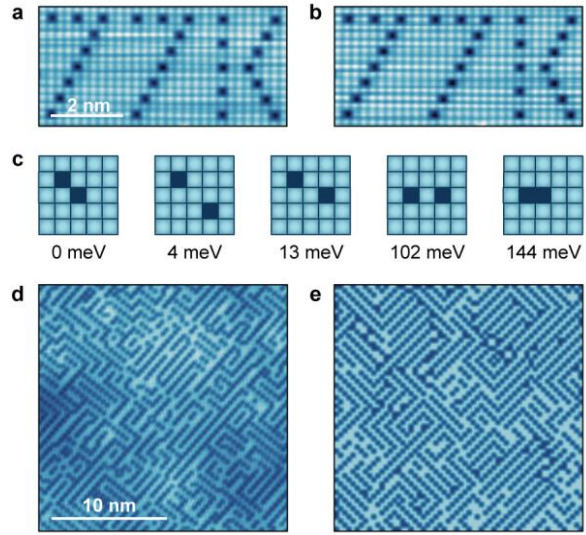

**Fig. 4. Thermal stability and vacancy-vacancy interactions. a.** STM image ($I$ = 300 pA, $V$ = +200 mV) of arranged vacancies measured at 77.5 K. **b.** Same as (a), but taken 44 hours later. **c.** DFT calculations on the energetics of pairs of vacancies, indicating that orientations along the (1,1) direction are strongly favoured over orientations along the (1,0) direction. **d.** STM image ($I$ = 500 pA, $V$ = +500 mV, $T$ = 1.5 K) taken at a vacancy coverage $x$ = 0.309, showing a stripe pattern. **e.** Monte Carlo simulation for $T$ = 100 K of a two-dimensional vacancy gas with $x$ = 1/3 and vacancy-vacancy interactions as shown in (c). Cell size 60×60, result shown after 400,000 steps starting from a random distribution.

Supplementary information for:

# A kilobyte rewritable atomic memory

F. E. Kalff[1], M. P. Rebergen[1], E. Fahrenfort[1], J. Girovsky[1], R. Toskovic[1], J. L. Lado[2], J. Fernández-Rossier[2,3] and A. F. Otte[1]*

[1]Department of Quantum Nanoscience, Kavli Institute of Nanoscience, Delft University of Technology, Lorentzweg 1, 2628 CJ Delft, The Netherlands.

[2]International Iberian Nanotechnology Laboratory (INL), Avenida Mestre José Veiga, 4715-310 Braga, Portugal.

[3]Departamento de Física Aplicada, Universidad de Alicante, San Vicente del Raspeig, 03690 Spain.

*To whom correspondence should be addressed. E-mail: a.f.otte@tudelft.nl

**Autonomous vacancy manipulation**

The procedure for the automatic assembly of vacancy arrays works as follows. First, a marker is built at the top left of the area designated for a 64-bit data block. In order to define the scan frame, the STM tip locks onto one of the vacancies in the marker, using an atom tracker sequence. Provided that the scan angle and piezo calibrations are fixed, only one marker is needed to fully define the scan frame.

After completing a scan, image recognition is used to identify the positions of all the vacancies in the scan. The initial configuration is compared to the desired final configuration, and each vacancy is assigned to a final position by means of the Munkres algorithm[1], attempting to minimize the total distance to be traversed. Vacancies are guided to their destinations using an A* pathfinding algorithm[2], which prevents vacancies from running into each other or forming dimers.

The computer program keeps a list of assignments that still need to be completed, and sends commands to the STM accordingly. Commands include "move tip to vacancy at <location>" and "move vacancy into <direction>". The assignments are based on the current configuration of vacancies as tracked by the program. The STM sends the outcome of each task back to the program. If the STM reports that a vacancy moved in the wrong direction, assignments are recalculated based on the new configuration.

A particular difficulty in the guiding process is to prevent vacancies from blocking each other so that their final position cannot be reached. For this situation we use the following procedure. We identify the vacancies that form a blockage to a specific final position and sort these in order of increasing distance to this position. Next, we move the first vacancy on the list to the final position, the second vacancy to the former position of the first vacancy, and so forth.

**DFT calculations**

Density functional theory calculations were carried out with the Quantum Espresso package[3], using the PBE exchange correlation functional and PAW pseudopotentials. Relaxation of a unit cell with one Cl and four Cu layers showed that the top layer does not suffer strong deformations. This result did not show an important dependence on the number of Cu layers, remaining unchanged even with only two Cu layers. The structural changes in the presence of a Cl vacancy were calculated in a 3×3 super-cell, allowing full relaxation of the system. The main changes appeared in the position of the Cl atoms, which were pushed towards the site with the missing Cl atom. In comparison, the position of Cu barely changed.

The lowest energy path for a vacancy, shown in Fig. 1b, was calculated in the 3×3 unit cell, moving the position of one of the Cl atoms towards the vacancy, and calculating its energy in every step. In this calculation we constrained the position of all the Cl atoms except the mobile one. The mobile atom had the position in the direction of the vacancy fixed, whereas the other two coordinates were allowed to relax. The path followed by the Cl was above the Cu, avoiding becoming closer to other Cl atoms. The energetics of the path does not depend strongly on whether the other atoms were allowed to fully relax or not.

For the calculation of the vacancy-vacancy interaction, 5×5 supercells with two Cl vacancies were chosen. The different arrangements of vacancies were calculated, allowing all the Cl atoms to relax. It was obtained that the vacancies have a first neighbour repulsive interaction, preferring a diagonal arrangement. Qualitative results do not depend on whether the Cu atoms were also allowed to relax. The same calculation with a 4×4 unit cell gave similar results.

**Monte Carlo simulations**

From the energetics of the DFT calculation for the vacancy-vacancy interaction, we can build a classical lattice gas model defined on the square lattice $S$. In this model, each site can be either full or empty. For a given configuration, the energy of the system is

$$\mathcal{U} = \sum_{\mathbf{r},\mathbf{r}' \in S} f(\mathbf{r}-\mathbf{r}') n_\mathbf{r} n_{\mathbf{r}'}$$

where $n_\mathbf{r}$ is the occupation of the site $\mathbf{r}$ (1 for vacancy, 0 for filled Cl site). The interaction $f(\mathbf{r},\mathbf{r}')$ affects every pair of vacancies, and its value is taken from the energetics obtained by DFT:

$$f(\mathbf{r}-\mathbf{r}') = \begin{cases} 0.144 & \text{if} \quad \mathbf{r}-\mathbf{r}' \in \{(1,0),(0,1),(-1,0),(0,-1)\} \\ 0.102 & \text{if} \quad \mathbf{r}-\mathbf{r}' \in \{(2,0),(0,2),(-2,0),(0,-2)\} \\ 0.013 & \text{if} \quad \mathbf{r}-\mathbf{r}' \in \{(2,1),(1,2),(-2,1),(1,-2),(2,-1), \\ & \qquad\qquad (-1,2),(-2,-1),(-1,-2)\} \\ 0.004 & \text{if} \quad \mathbf{r}-\mathbf{r}' \in \{(2,2),(2,-2),(-2,2),(-2,-2)\} \\ 0 & \text{otherwise} \end{cases}$$

where the energies are in eV. Using the standard Metropolis update algorithm, we carry out Monte Carlo simulations within the canonical ensemble, i.e., with a fixed total number of vacancies, at a given temperature $T$. The temperature enters in the Metropolis update as the tolerance for accepting a new configuration with higher energy. Starting from an initial random configuration, at every step the update algorithm attempts to move a random vacancy to a random neighbouring filled site. For $T \to 0$, only new configurations that lower the energy are accepted, so that a local energy minimum is finally reached. For $T \to \infty$, any new configuration is accepted, leading to a fluctuating and disordered state.

Starting from a randomly generated configuration, the Monte Carlo evolution drives the system towards a local Free energy minimum. The system develops different domains with stripes in different directions, as observed in the experiment. For the calculations in the main text, a relaxation of 400,000 steps was carried out. Finally, it is worth to note that the results do not change qualitatively if a simpler interaction is assumed:

$$f(\mathbf{r} - \mathbf{r}') = \begin{cases} 0.144 & \text{if} \quad \mathbf{r} - \mathbf{r}' \in \{(1,0), (0,1), (-1,0), (0,-1)\} \\ 0 & \text{otherwise} \end{cases}$$

**Supplementary Movie S1.** Autonomous assembly of one 64-bit block. Real-time screen capture taken during automated vacancy manipulation without human interaction. At the beginning of the movie, all shown windows are briefly introduced. The movie starts at the moment a scan of the area has been taken and interpreted by the image recognition program. Three atomic markers, which were built as part of previous blocks, are already in place. The movie shows 264 hops, one of which is in an unintended direction.

Movie link:
https://www.dropbox.com/s/u80zorua38iwwfo/Vacancy%20manipulation.mp4?dl=0

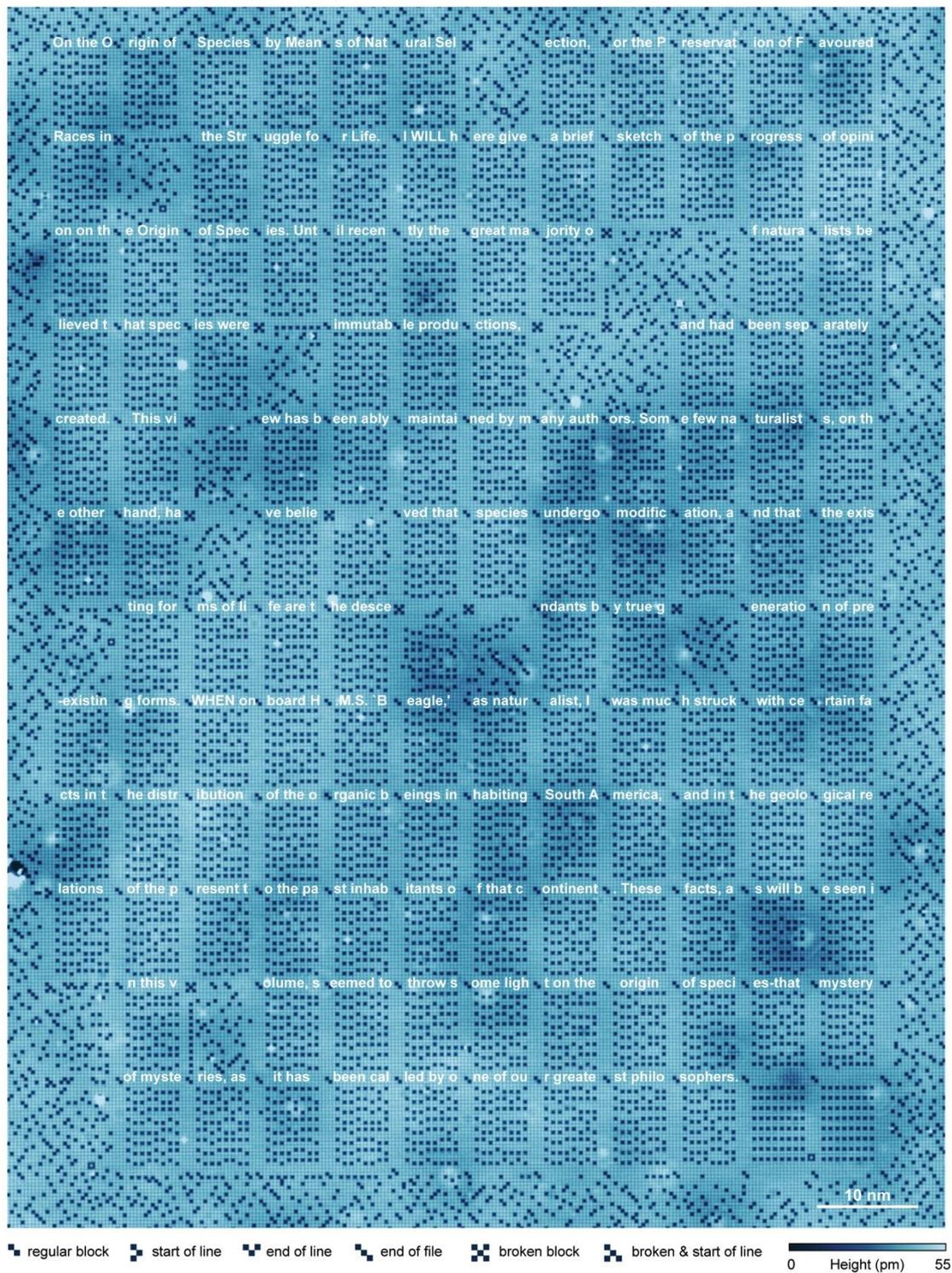

**Fig. S1.** STM image (96 nm × 126 nm, $I$ = 2.00 nA, $V$ = +500 mV, $T$ = 1.5 K) of the same 1016 byte memory as shown in Fig. 3, written to a different text. The presented text is an excerpt from the preface and the introduction of *On the Origin of Species*, by Charles Darwin (slightly adapted to fit the memory).

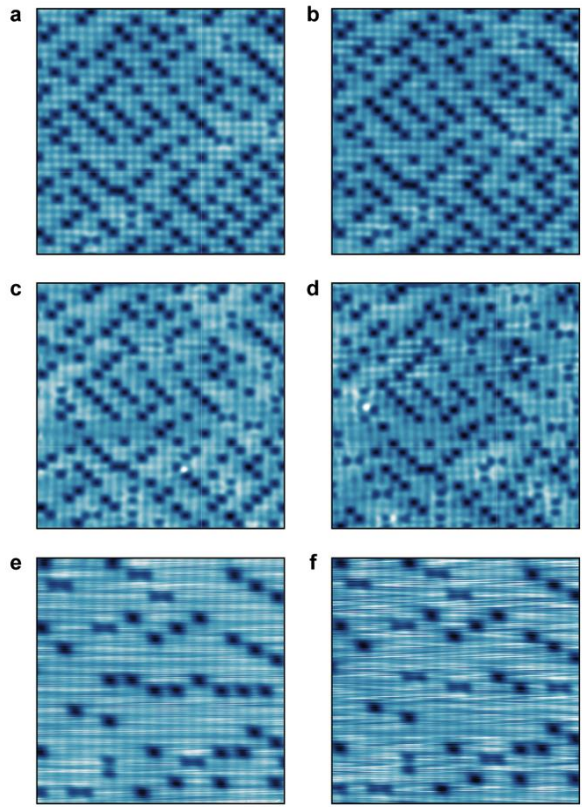

**Fig. S2. a.** STM image of vacancies in a natural configuration, measured at 77.3 K. The image shows five 'bowtie'-shaped objects, the exact atomic structure of which is unknown. **b.** Same as (a), but measured 19 hours later. No vacancy has moved. **c,d.** STM images of the same area as (a) and (b), but measured at 78.2 K and 78.4 K, respectively, taken only minutes after (b). The number of bowtie objects is found to increase with increasing temperature. **e.** STM image of vacancies arranged to form a byte, measured at 77.5 K. **f.** Same as (e), but measured 50 minutes later. Several vacancies positioned in the (2,0) configuration have changed into bowtie objects.